\documentclass[6 pt,conference]{IEEEtran}
\usepackage{amsmath}
\usepackage{booktabs} % For formal tables
\usepackage[lined,boxed]{algorithm2e}
\usepackage{multirow}
\usepackage[font=small,skip=1pt]{caption}
\usepackage{subcaption}
   \usepackage[pdftex]{graphicx}     
   \graphicspath{{../pdf/}{../jpeg/}{./image/}}    
 \DeclareGraphicsExtensions{.pdf,.jpeg,.png,.jpg}

\IEEEoverridecommandlockouts                              % This command is only
\DeclareMathOperator{\nps}{\textit{NPS}\textsuperscript{\textregistered}}

\DeclareMathOperator{\sps}{\textit{SPS}}

\DeclareMathOperator{\base}{\textit{SPS-R}}
\title{Social Promoter Score ($\sps$) and Review Network: A Method and a Tool for Predicting Financial Health of an Online Shopping Brand}

\author{ Supriyo Mandal$^{1}$, Abyayananda Maiti$^{2}$ 
\\Department of Computer Science and Engineering
\\Indian Institute of Technology Patna, 
\\Patna, Bihar, India 801103
\\email: $\lbrace$ $^{1}$supriyo.pcs17, $^{2}$abyaym$\rbrace$@iitp.ac.in}

\begin{document}

\maketitle
\thispagestyle{empty}
\pagestyle{empty}

%%%%%%%%%%%%%%%%%%%%%%%%%%%%%%%%%%%%%%%%%%%%%%%%%%%%%%%%%%%%%%%%%%%%%%%%%%%%%%%%

\begin{abstract}
Conventional way of summarizing customers' ratings or sentiment of reviews on products of an online shopping brand are not sufficient to evaluate the financial health of that brand. 
It overlooks the social standing and influence of individual customers. 
In this paper, we have  proposed a tool named as Review Network for measuring influence of customers in online merchandise sites like Amazon.com.
Using this measured influence, we have proposed a method that evaluates loyalty of customers of a brand based on their ratings and sentiments of their reviews collected from online merchandise sites.
Review network of a brand is built from all the reviews of all the products from that brand where nodes are customers and an edge is created if a customer becomes a potential reader of a review written by another customer. 
Centrality of a customer in that review network represents her influence. 
Our proposed method named as Social Promoter Score ($\sps$) combines loyalty and centrality of all customers of a brand.
We have compared our method with a base line approach named as $\base$ based on the concept of $\nps$\cite{NPS}. 
We have applied $\sps$ on Amazon.com review data set of some well known brands, i.e., Nokia, Samsung, Philips, Canon, Lenovo etc.
Results show that $\sps$ predicts financial health of a brand in terms of future sales much better than $\base$. 
We have noticed that in general effects of $\sps$ reflect on product sales in one to five months. 
%\footnote{This is an abstract footnote}
\end{abstract}

%%%%%%%%%%%%%%%%%%%%%%%%%%%%%%%%%%%%%%%%%%%%%%%%%%%%%%%%%%%%%%%%%%%%%%%%%%%%%%%%
\section{Introduction}\label{scc:intro}

%The \textit{proceedings} are the records of a conference.\footnote{This
% is a footnote} 
%

Predicting financial health is always one of the top prioritized tasks of any company's or brand's higher managements.
It helps a manager to keep track of a company's performance in advance to identify any future risk in company's growth profile.
In extreme cases financial health prediction warns about unwanted phenomenon like bankruptcy and failure.
Besides, regulatory agencies can use this for monitoring company's future.
In general financial health prediction gives a good indicator of future revenue growth of the company and helps to build business and marketing policies.
In this paper we use the terms \textit{brand} and \textit{company} synonymously.

Prediction of corporate financial health of a company from its financial data is an important research topic for long time.
Different machine learning techniques have been used to analyze finance data for financial health prediction~\cite{Rafiei2011}.
But these analyses depend only on past and present performance of the company and overlook the image of the company (brand's image) among its present and potential customers.
It has been shown in many studies that brand awareness positively affects company's performance~\cite{huang2014,oliver2014,aaker2012}.
Customer satisfaction and expectation which are both outcomes and causes of company's image affects the profitability~\cite{anderson1994}. 
Retaining a customer or make a customer loyal to a brand further boost the company's growth~\cite{reichheld1990}.
Several studies report a strong positive relation between customer loyalty and customer satisfaction~\cite{anderson1993}.

In this context we observed two different conceptualizations of customer satisfaction: transaction-specific and cumulative~\cite{boulding1993}. 
Transaction-specific satisfaction is more customer-centric concept where satisfaction is measured on individual customer's purchasing experience.
On the other hand, cumulative satisfaction is based on aggregate satisfaction level of all customers over time~\cite{johnson1991}.
In general, cumulative satisfaction is regarded as more fundamental indicator of the company's performance~\cite{anderson1994}.

In business world most popular metric to gauge the loyalty of a brand's customer relationships is Net Promoter Score ($\nps$). 
$\nps$ is registered trademark of Frederick Reichheld, Bain \& Company, and Satmetrix Systems. 
This concept was published in \textit{Harvard Business Review} in 2003~\cite{NPS}.
However, it is still being widely adopted with more than two thirds of Fortune 1000 companies~\cite{kaplan2016}.
This concept suggests that of all the customer survey metrics a company can track, one is the most important in terms of its relationship with brand's financial performance.
It asks  just one question: ``How likely is it that you would recommend our company/product/service to a friend or colleague?" in 0 to 10 scale.
A manual survey of customer feedback is done with that one question about customer satisfaction. 
$\nps$ identifies a customer as promoter if she gives 9 or 10 rating. 
Who gives response with 7 or 8 are called as  a passive customer and responses with 0 to 6 rating is known as a detractors. 
$\nps$ is defined as (\% of promoters -  \% of detractors). 
A positive $\nps$ is considered to be good and a $\nps$ of 50+ is treated as excellent. 
At the initial stage it is evaluated from the survey of word-of-mouth ($WOM$) technique~\cite{WOM}.

Although Net Promoter Score has gained popularity among corporate executives, it has attracted a fare share of criticism from academia and market research circles~\cite{grisaffe2007}.
The main drawback of $\nps$ is that it is a cumulative satisfaction measure with only one parameter of recommendation.
It overlooks the satisfaction level of individual customers~\cite{raassens2017} and the effects of good marketing strategies of the brand. 
Furthermore, giving a good recommendation to customer feedback survey does not guarantee a positive word-of-mouth marketing by that customer in reality.
In this paper we present a framework for measuring customer loyalty which tries to rectify the above mentions drawbacks based on online reviews of products of a brand in online merchandise site like Amazon.com. 

The world is changing rapidly and as the Internet grows, online shopping has played a vital role for purchasing products. 
In many cases, due to its global nature, lower cost, fast delivery, time saving and $24/7$ hours availability online merchandise platforms like Amazon, eBay are the preferred one for acquiring products.  
In 21$^{st}$ century the popularity of online merchandise systems is increasing rapidly. 
Online shopping results in explosive growth in instant customer surveys.
There is a feedback section in every online merchandise system where customers submit their feedback about their purchased products.
Now a days companies are more eager for client feedback to improve products and services.
The increasing focus on data, the ease of reaching customers via online platforms, and the growing conviction that by rating a product customers gain a stake in it makes the customers members of that product's ``community".
According to recent studies those under 35 have grown used to the above mentioned approach and consult many online reviews before making a purchase~\cite{kaplan2016}.

Many methods have recently been proposed\cite{brand,sentiment1,sentiment2} which classify a review as positive or negative. 
In \cite {review1,review2} authors have analyzed sentiment of customers' online review text using some machine learning approaches.  
They have proposed an online version of $\nps$ which makes the classification of promoters and detractors based on the sentiment score of the reviews.
In some of the online merchandise sites like Amazon.com reviews have a collaborative mechanism to build its reliability by means of helpfulness score.

As mentioned earlier, traditional cumulative approach like $\nps$ gives uniform importance to all customers who are writing reviews in online platforms.
In reality, different customers have different capability of promoting the products depending on their online word-of-mouth (eWOM) power~\cite{raassens2017}.  
Customers' influence and connectivity in their social networks and/or commerce networks determine their eWOM power.

Now a days people are very much attached with their social network friends. 
Whatever they like they share it in their social networks. 
Some methods  are proposed in~\cite{{socialmax,socialmax1,socialmax2,socialmax3}} for monitoring user interactions and generating proactive responses with in a social media environment and calculate real time customers' satisfaction level. 
The authors calculate customers' loyalty or customers' rank based on social media interactions.
Customers' influence and connectivity can be evaluated from their centrality values in their respective social networks.
Sometimes it is difficult to figure out the influence of customers for a certain product of a company in social networks due to limited data availability in public domain.
Customers' commerce networks can give us more accuracy on the influence they incur on other customers.

Commerce networks are built from customers, organizations and activities of  buying and selling different products. 
Over the past decades researchers have extensively studied different offline commerce networks and try to propose methods to predict commerce behaviors based on customers' satisfaction level with the products of companies~\cite{shen2011}. 
Traditionally, from offline surveys and questionnaires or simulated from theoretical models, commerce networks are obtained. 
In this paper we have proposed a new type of commerce network, namely Review Network. 
Review network of an online shopping brand is built from all the reviews of all the products from that brand where nodes are customers and an edge is created if a customer becomes a potential reader of a review written by another customer.
In this paper we use review networks based on Amazon.com review data set as a tool to apply our proposed methodology, namely Social Promoter Score ($\sps$). 
$\sps$ is a method that calculates reputation score of a customer based on her rating and sentiment of her reviews and how many potential customers from her social networks or review network could be influenced or detracted by her when she gives feedback regarding products in the merchandise cite or shares her posts in  her social media.
Aggregation of reputation scores from all customers gives $\sps$ of the brand.

Based on customers' loyalty a brand identifies influential customers who are crucial for promoting its products. 
Several methods are implemented in~\cite{fackorloyal} that track loyal customers or fake customers and identify a key influencer in a social media environment for enterprise marketing utilizing topic modeling and social diffusion analysis.
% 
%Loyal customers with high centrality score and detractor are identified by the company more accurately and depending on social standings company can decide different incentive schemes for specially influential customers and  establish strategy to encourage the detractors to purchase products again by enhancing the quality of products or by more aggressive advertisements/ incentives or both. 
%
%Some methods are described in \cite{advertise,advertise1}  where advertisers wish to deliver some offer related advertisement or incentives to the selected customers based on their loyalty score.
%
A large volume of research on networks like friendship network, customers' review network, social media based network such as, for example,  blogs and microblogs (e.g. Twitter),  social networking sites (e.g. Facebook) and virtual social worlds (e.g. Second Life) has been devoted to the concept of centrality such as degree centrality, eigenvector centrality\cite{eigenvector}, katz centrality, betweenness centrality\cite{betweenness}, closeness centrality\cite{closeness} etc. to identify influential nodes in those  networks. 

The major contributions of this paper are summarized next:
\begin{itemize} 
\item We consider helpfulness score for each customer to a product that is used to validate the reliability of the particular customer's product review and rating.
\item Loyalty score of each customer is evaluated from customer's helpfulness score and customer's rating score or sentiment of her review text on her purchased products.
\item We evaluate the influence of customer from her centrality score in her social or commerce network. We conceptualize a new type of commerce network, namely review network.
\item Reputation score of each customer is evaluated from her centrality score and loyalty score. Aggregation of all customers' reputation score is the $\sps$ of the brand.
 \item We have shown that $\sps$ of a brand can predict its financial health better than a base line approach $\base$.   
\end{itemize}

The rest of this paper is organized as follows. 
In next section we define the problem. 
In Sec.~\ref{sec:sps} we explain our proposed method and define the baseline approach that we consider.
The concept of Review Network is presented in Sec.~\ref{sec:review_net}.
In Sec.~\ref{sec:result} empirical evaluation of $\sps$ is given with Amazon.com review data set for some well known brands.
Lastly, we give conclusion and future work.

\section{Problem Formulation} \label{sec:problem}
In business world,  customer satisfaction and loyalty of a brand are evaluated based on customers' ratings and review texts.
By calculating total number of promoters and detractors, loyalty score of a brand is evaluated. 
The company fix a rating scale to define who are promoters and who are detractors. 
It is not sufficient to evaluate brand's popularity to predict its future financial market growth. 
Brand's financial health depends on  customers' loyalty score and also on how many customers are influenced (positively or negatively) by them.
We consider this concept and evaluate brand's loyalty and consequently its financial health. 

Let assume a brand is registered in an online merchandise site like Amazon.com.
The brand has $n$ number of  products $p_1, p_2, ..., p_n$ on sale.
There are $m$ customers/buyers $u_1, u_2, ..., u_m$ who are registered users in that merchandise site.
Customer $u_j$ writes review $r_{ij}$ for product $p_i$ after purchasing it online.
Similarly the ratings given by $u_j$ for $p_i$ is denoted by $q_{ij}$.
Our problem is to predict financial health of the brand from the reviews/ratings and the structures of the participating networks.

\section{Social Promoter Score ($\sps$)} \label{sec:sps}
Customer Relationship Management ($ CRM$) evaluates customers' loyalty based on their reviews and ratings on a brand or brand's products that help to analyze customers' relationship with the company and  with the goal of improving business relationships with customers, assisting in customer retention and driving financial growth. 
Now a days it has become more important how to manage a long-term customer relationship because it can assure the company to increase income in spite of steep competition with rival brands. 
Therefore a healthy customer relationship is an important means of proving the brand with competitive edge and maximizing brand's sales. 

Our proposed  method  evaluates  each customer's loyalty score to the brand based on her ratings and reviews from online  (e-commerce) merchandise platforms. 
Centrality score of each customer is also calculated based on how many customers are influenced by her reviews and ratings to purchase the same product. 
Loyalty score and centrality score both will be  effective to predict  financial health of the brand.

\subsection{Online Merchandise System} \label{subsec:online_merh}
Online merchandise sites (such as Amazon, eBay) save time for the customers and these are available $24/7$ hours. 
%
%For these reasons now a days popularity of online merchandise sites are increasing rapidly. 
%
Sometimes third party sellers manage the product delivery.
Customers order different products of a particular brand through online merchandises' website or mobile application. 
The orders go to official cloud server. 
The company disburses products  to the third-party sellers. 
The third parties deliver the products to the particular customers. 
%
%Admin  has permission to access all the credential information from official server. 
%
%

The third parties deliver the products with in the given time period.  
After receiving the products customers submit feedback in terms of rating and review to the server. 
Customer service center also collects customers' feedback through telephonic interactions. 
Some customers may also share social posts regarding their purchased products on their social network platforms. 
All data are shared with company database  and analytic management  investigates on  company's database for various business intelligence.

\subsection{Customer Helpfulness Score}\label{subsec:helpful}
After purchasing product, customer gives ratings and reviews, associated with the products in the online merchandise site. 
Before purchasing that same product other customers are expected to read the previous reviews regarding that particular product.
In most of the merchandise sites after each review it asks the question,  ``Was this review helpful to you? (Answer Yes/No)''. 
In this case, the answer to the question corresponds to the feedback on the review. 
``Yes'' answer indicates that the review is helpful to the customer. 
``No'' answer indicates that the reviews are not proper or not truthful.
This helpfulness data can be used to validate the reliability of the particular customer's product review. 
We define helpfulness score $H_{ij}$ of the review given by customer $u_{j}$ for product $p_{i}$ as follows: 
\begin{equation}\label{eq:help_score}
H_{ij} = \frac{x_{ij}}{y_{ij}}
\end{equation} 
where $x_{ij}$ is the number of customers who marked the review given by customer $u_{j}$ for product $p_{i}$ helpful and $y_{ij}$ is the total number of customers who have answered that question (both yes and no).

\subsection{Customer Loyalty Score}\label{subsec:loyalty}
%
%After purchasing  products, customers give ratings and write reviews about those products. 
% 
We  calculate loyalty score of customers from rating and sentiment score of their reviews.
In this paper both ratings and sentiment scores are assumed to be integral.
However, real valued ratings and sentiment do not change the methodology.
We scale both ratings and sentiment scores from $-2$ to $+2$ i.e. $(-2, -1, 0, +1, +2)$. 
For example, Amazon.com ratings span from $1$ to $5$.
So in this case a rating of $5$ would be scaled down to $+2$ and $1$ would be $-2$.

We evaluate sentiment score from customers' reviews using one or more machine learning algorithms. 
This process is also known as sentiment analysis.
If sentiment score is positive then customer is happy with her purchased product and if sentiment score is negative then customer is not satisfied with her purchased product.
In this case also we have five level of sentiments.

We make a realistic assumption that if a customer rates and writes review for a purchased product, then rating score and sentiment score would be same because both rating score and sentiment score are about the experience of the customer for that particular product.  
In rest of the paper we use the term sentiment score to represent both rating and sentiment score.
If both rating and sentiment of review are available we can take any one of those two scores as sentiment score.
We denote sentiment score as $R_{ij}$ for rating $q_{ij}$ or review $r_{ij}$.
If rating $q_{ij}$ or review $r_{ij}$ do not exist then $R_{ij} = 0$.
Loyalty score of customer $u_j$ to product $p_i$ is denoted by $L_{ij}$
This is calculated  from  customer's sentiment score multiplied by helpfulness score.
\begin{equation}\label{eq:loyalty_score1}
L_{ij} = \frac{x_{ij}}{y_{ij} }R_{ij}
\end{equation}       
If helpfulness score is not available then loyalty score becomes
\begin{equation}\label{eq:loyalty_score2}
L_{ij} = R_{ij}
\end{equation} 

\subsection{Customer Centrality Score}\label{subsec:centrality}
We use centrality score of a customer in her social or commerce network to evaluate her influence of promoting certain product.
If a customer buy a product, we filter out the sub-network from the social and commerce network related to that purchase only.
For example if customer $u_j$ purchases product $p_i$ of a particular brand and post a message in her social network about that then we will consider the portion of her social network with $u_j$ and her friends who can read that post.
So for each pair of $u_j$ and $p_i$ we have a sub-network.
Centrality score of $u_j$ in that sub-network is denoted as $D_{ij}$ which is normalized by the maximum centrality among all the customers of the brand.

There are many different kinds of centrality measures such as degree centrality, eigenvector centrality, katz centrality etc. %
It depends on the objective of the company's management which centrality measure we would consider.
In this paper we use degree centrality to identify the most important or central nodes in networks.

\subsection{Calculation of $\sps$}\label{subsec:sps_calc}
We calculate reputation score of each customer based on her centrality score and loyalty score. 
Further we aggregate reputation scores of all customers and compute Social Promoter Score (SPS), associated with a brand.  
Loyalty score of a customer for a particular product is multiplied by centrality score of the customer for that product to get her reputation score for that product. 
We evaluate reputation score of the customer for all of her purchased products.
If helpfulness score is available then we add an extra factor based on $x_{ij}$ in the equation.
The expression of $\sps$ is given as:

\begin{equation}\label{eq:sps}
\sps = \sum_{j=1}^{m} \sum_{i=1}^{n}  \left(D_{ij}+x_{ij}\right) L_{ij}.
\end{equation}

In the above equation we put $x_{ij}$ because helpful reviews or messages have profound impression on potential customers on selecting some products.
Most customer in online merchandise sites rely on reliable reviews to make a decision. 
Higher value of $x_{ij}$ indicates that collaboratively more number of customers feel that the decision of the reviewer is correct.
Substituting the expression of $L_{ij}$ from Eq.~\ref{eq:loyalty_score1} in Eq.~\ref{eq:sps} gives
\begin{equation}\label{eq:sps1}
\sps = \sum_{j=1}^{m} \sum_{i=1}^{n}  \left(D_{ij}+x_{ij}\right) \frac{x_{ij}}{y_{ij}} R_{ij}.
\end{equation}

\begin{algorithm}
\SetKwData{Left}{left}\SetKwData{This}{this}\SetKwData{Up}{up}
\SetKwFunction{Union}{Union}\SetKwFunction{FindCompress}{FindCompress}
\SetKwInOut{Input}{input}\SetKwInOut{Output}{output}
\Input{Set of normalized centrality score $D_{ij}$, Set of sentiment score $R_{ij}$, Set of helpfulness score $\frac{x_{ij}}{y_{ij}}$, and set of helpful customers $x_{ij}$.}
\Output{$\sps$ }
   \emph{Initialization: SPS :=0 ; } \\
    $n$ is the total number of products and $m$ is the total number of customers; \\ 
\If {helpfulness score is available}{
\For{$j\leftarrow 1$ \KwTo $m$ } {
\For{$i\leftarrow 1$ \KwTo $n$}{ 
  
$\sps$ = $\sps$ + ($D_{ij} + x_{ij}$)$\frac{x_{ij}}{y_{ij} } R_{ij}$ \;

 }

}
%SPS = $S_{ij}$\;
}
\Else{
\For{$j\leftarrow 1$ \KwTo $m$}{
\For{$i\leftarrow 1$ \KwTo $n$}{

$\sps$ = $\sps$ + $D_{ij}  R_{ij}$ \;

 }

}
 
}
\caption{Social Promoter Score (SPS)}\label{algo:sps}
\end{algorithm}\DecMargin{1em}
If helpfulness score is not available then The $\sps$ is expressed as:        
\begin{equation}\label{eq:sps2}
\sps = \sum_{j=1}^{m} \sum_{i=1}^{n}  D_{ij} R_{ij}.
\end{equation}

In algorithm~\ref{algo:sps} we have shown the step by step process of $\sps$ calculation that is already discussed in Eq.~\ref{eq:help_score} to Eq.~\ref{eq:sps2}.
From $\sps$ value brands can predict more accurately their future financial market growth. 
A higher value of increment of $\sps$ than a predefined threshold in between certain time period means the degree of customer loyalty to the brand is good and influential customers are happy with the products of that brand. 
As a result company's market growth will be positive.
Please note that threshold value always varies for different brands that will be decided by the company. 
On the other hand low or negative increment of $\sps$ means the company has to change its strategies towards the influential customers either by enhancing the quality of products or by more aggressive advertisements/incentives or both. 
There will be a latency period which gives the time required for showing the effect of $\sps$ on actual growth of the brand.
This latency period depends on data, season, brand offers, marketing strategy etc. 

\subsection{Baseline $\base$}\label{subsec:baseline}
To evaluate the efficiency of our method $\sps$ we compare it with an alternate version of the standard measure $\nps$\cite{NPS}.
As mentioned earlier, $\nps$  can be formulated from ratings and sentiment of reviews.
We propose a baseline approach named $\base$ based on the concept of $\nps$ where centrality and helpfulness values are ignored.

From the review data set of the online merchandise site we collect sentiment score $R_{ij}$ of all customers of all products of a company in between certain time period. 
Then we identify the promoters and detractors set in between that period. 
If  $\base$ is required for a particular month or particular year or certain time period, then we identify the promoters and detractors set of the particular month or particular year or certain time period respectively.  
In algorithm \ref{algo:base} we fix a predetermined lower bound on sentiment score ($B_l$) for promoter and and upper bound of sentiment score ($B_u$) for detractor.
Accordingly we evaluate the set of promoters and detractors of a brand. 
Please note that same customer may be counted as promoter for one product and detractor for another product depending on her reviews/ratings on those two products. 
$\base$ is equal to \% number of promoters - \% number of detractors. 
%
%Number of promoters and number of detractors are scaled by total number of reviews in between that period. 
%
There will be some reviews which give passive customers who do not fall in to the either of those two sets.

\begin{algorithm}
%\DecMargin{2em}
\SetKwData{Left}{left}\SetKwData{This}{this}\SetKwData{Up}{up}
\SetKwFunction{Union}{Union}\SetKwFunction{FindCompress}{FindCompress}
\SetKwInOut{Input}{input}\SetKwInOut{Output}{output}
\Input{Set of sentiment score $R_{ij}$ evaluated from review data set in between certain time period $T$}
\Output{$\base$ in time period $T$}

   \emph{Initialization: number of promoters ($p_{r}$):=0; \\ number of passives ($pa_{r}$):=0; \\number of detractors ($d_{r}$) :=0; }\\
$m$ is total number of customers who wrote reviews on $n$ products in between time period $T$; \\      

\For{$j\leftarrow 1$ \KwTo $m$}{
\For{$i\leftarrow 1$ \KwTo $n$}{

\If {$u_j$ gives rating/review on $p_i$ during $T$}{
\If {($R_{ij}$ $\geq$ B$_{l}$)}
  {
    $p_{r}$++  \;
   } 
   \ElseIf {($R_{ij}$ $\leq$ $B_{u}$)}
 {
    $d_{r}$++\;   
  } 
  \Else 
  {
   $pa_{r}$++\; 
   }
   
   }    
}
}

$A$= $p_{r}$+$pa_{r}$+$d_{r}$; /* $A$ is total number of reviews in between certain time period */\\
$\base$ = $\dfrac{p_{r}}{A}$*100\% - $\dfrac{d_{r}}{A}$*100\% ;

\caption{$\base$ Calculation for time period $T$}\label{algo:base}
\end{algorithm}

A basic difference between $\sps$ and $\base$ is that $\sps$ is a accumulative score which grows as the underlying network grows whereas $\base$ is a periodic measure like $\nps$.
Calculation of $\base$ for current month does not depend on previous months.
For this reason we always compare increments of $\sps$ with $\base$.

\section{ Review Network}\label{sec:review_net}
\begin{figure}[t]
\centering
\includegraphics[scale=.451]{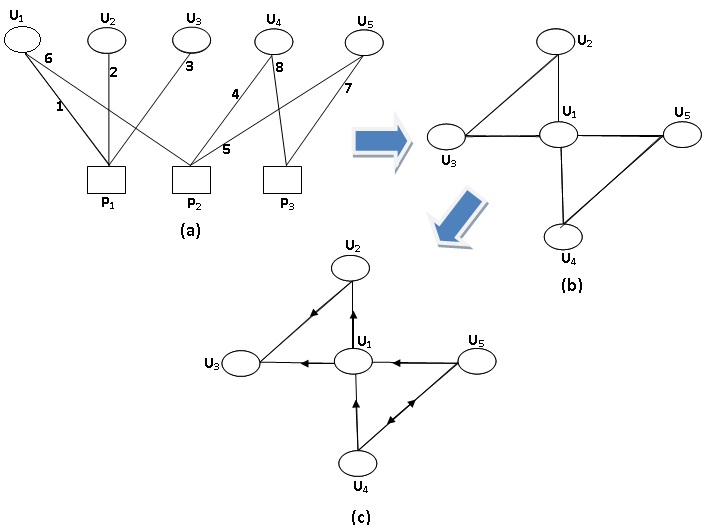}
\caption{Step by step building of Review Network. (a) Bipartite network between products and customers. An edge denotes a review written by a customer on a product. (b) One mode projection on customers (c) Directed network based on time stamps of reviews. In a directed edge the source node influences the end node.}
\label{fig:review_net}
\end{figure}
Since most of the social network data are private, it is hard to evaluate the influence a customer has in her social network related to certain online purchase.
If that customer writes a review on that purchase product in the merchandise site then the influence of that purchase can be measured as the review data are publicly available. 
When we are going to purchase any product from online merchandise sites, we read previous customers' reviews which are related to that particular product.
If any customer read those reviews and purchase that product and write another review on that product then we can infer that the last reviewer is influenced by all of the previous reviewers.
We can build a network of reviewers where a directed edge from a previous reviewer to the last reviewer denotes influence.
So, in that network outgoing degree of a reviewer node specifies the total influence of that node.
We name this network as \textit{Review Network}.      

Fig.~\ref{fig:review_net} shows the step by step process of building review network.
Fig.~\ref{fig:review_net}(a) depicts a bipartite network presenting the review data set.
Two sets of nodes in that bipartite network are product set and customer set.
If customer $u_j$ writes a review on product $p_i$ then there will be an edge between them. 
Essentially each edge represents a distinct review.
Notice that, each review has a time stamp of its creation. 
We identify each edge by unique number that indicates logical time stamp of edge creation. 
Please note that here we have not specified original time stamps. 
For depiction purpose we have assumed some time direction in Fig.~\ref{fig:review_net}(a). 
The edges between customer $u_1$ product $p_1$ and customer $u_2$  product $p_1$ are identified by time stamps $1$ and $2$ respectively that means customer $u_2$ purchases product $p_1$ after customer $u_1$. 

In Fig.~\ref{fig:review_net}(b), we create an undirected version of review network.
This is an one mode projection on customer set of the customer-product bipartite network.
An edge between two customers in the one mode projection means both of them have written review for the same product.
In Fig.~\ref{fig:review_net}(c) we covert simple edges to directed edges.
Here direction follows the time line. 
It denotes that who ever writes a review on a product later she has read all the previous reviews on that product.
So outgoing degree specifies the number of influenced customers.. 

In Fig.~\ref{fig:review_net}(a), $p_1$, $p_2$, $p_3$ are products of a brand and $u_1$, $u_2$, $u_3$, $u_4$, $u_5$ are customers have written reviews on one or more products from those three products.
In figure Fig.~\ref{fig:review_net}(b), users $u_1$, $u_2$ and $u_3$ are connected with each other because all of them have written reviews on product $p_1$.
In figure Fig.~\ref{fig:review_net}(c) there is a directed edge from $u_1$ to $u_2$ since $u_2$ wrote review on $p_1$ after $u_1$ wrote. 
In other words, $u_2$ read $u_1$'s review and influence by that review she purchased $p_1$ and then wrote another review on $p_1$.
Out-degree of $u_1$ is $2$ denotes that she influenced two customers to purchase some products.
So, centrality score of $u_1$ in this review network is $2$. 
In-degree of $u_1$ is $2$ means she got influenced by two reviews to buy some products.
In this model parallel edges are allowed.

Please note that, in Fig.~\ref{fig:review_net} we have presented a whole review network based on a set of customers and products.
However, in calculation of $\sps$, centrality score will be calculated on filtered sub-network based on each review.
So, if $u_1$ wrote reviews on both $p_1$ and $p_2$ and we calculate $D_{11}$, we do not consider how much influence $u_1$ has on the customers who read $r_{21}$.
Effectively, $D_{11}$ will be the number of reviewers who write reviews on $p_1$ after $r_{11}$ is written.  

\section{Experimental Results}\label{sec:result} 
In this section we present our experimental result on $\sps$.
We apply $\sps$ on review network generated from Amazon.com review data.
First we show that monthly increment of $\sps$ has a high positive correlation with monthly product sale of companies/brands listed in Amazon.com.
Then we show that in this regard $\sps$ is doing much better than the baseline approach $\base$.
Lastly we discuss some interesting observations from our experimental results.

\subsection{Data Statistics}\label{subsec:data}

The data used in our experiments, is collected from Amazon.com online review data set~\cite{he2016,mcauley2015}, associated with different brands sold in Amazon.com. 
This data set provides all the required information about each review to build a corresponding review network.
Here is a sample review entry in the data set: \\
\textit{\{\\
  ``reviewerID": ``A2SUAM1J3GNN3B",\\
  ``asin": ``0000013714",\\
  ``reviewerName": ``J. McDonald",\\
  ``helpful": [2, 3],\\
  ``reviewText": ``I bought this ..... Great purchase though!",\\
  ``overall": 5.0,\\
  ``summary": ``Heavenly Highway Hymns",\\
  ``unixReviewTime": 1252800000,\\
  ``reviewTime": ``09 13, 2009"\\
\}\\}
Here $``asin"$ means product id and  ``$overall"$ denotes rating value.
Helpfulness score is taken form ``$helpful"$ attribute and ``helpful": [2, 3] means two customer think that the review is helpful and one customer feels it is not helpful.
We identify the brand name from $``reviewText"$ field of the review.

We have taken only rating value to evaluate sentiment score.
In Amazon.com rating values are from $1$ to $5$.
We scale them in $-2$ to $+2$.
Positive rating score denotes promoter (lower bound $B_l = 1$), negative rating score denotes detractor (upper bound $B_u = -1$) and zero rating score denotes passive customer.

From this data set we build review networks and accordingly apply our proposed methodology $\sps$ on some popular brands such as Nokia, Samsung, Lenovo, Philips and Canon for determining the performance of those brands.
For comparing the results we also calculate $\base$ for each brand in between certain time periods (mostly monthly).
We have taken review data of those brands from 2002 to 2011.
Table~\ref{tab:stat} shows the number of unique customer id, unique product id and reviews for each company.
\begin{table} [hbtp]
%\addtolength{\tabcolsep}{.001pt}
\centering
\caption{Statistics of the Data set} 
%\footnotesize
\scalebox{.9}{
\begin{tabular}{|c c c c|}
\cline{1-4}
\multirow{1}{*}{\textbf{Company}} &\multirow{1}{*}{\textbf{\# reviews}} & \multirow{1}{*}{\textbf{\# customer}}\multirow{1}{*}& {\textbf{\# product}}  \\ \cline{1-4}
Nokia & 15186 & 10884 & 198  \\ %\cline{1-4}
Samsung & 47998 & 34298 & 1292  \\ %\cline{1-4}
Philips & 11561 & 10376 & 580  \\ %\cline{1-4}
Canon & 15526 & 9889 & 108  \\ %\cline{1-4}
Lenovo & 7271 & 6746& 205  \\ \cline{1-4}

\end{tabular} 
}
\label{tab:stat}
\end{table}
\vspace{.01 mm}

\subsection{Prediction of Financial Health}\label{subsec:health}
\renewcommand{\arraystretch}{1.4}
\begin{table*} [t]
\centering
\caption{Comparison between increment of $\sps$ with $\base$ in 2009 using Pearson correlation coefficient. In second column $\sps$ indicates increment of $\sps$. For each brand there are two rows for correlations between monthly $\sps$ increments and monthly product sales and  correlations between monthly $\base$ and monthly product sales over twelve months. Twelve months of product sales are given with different latency (offset) periods. $\sps$ and $\base$ are calculated for fixed twelve months (January 2009 to December 2009). Max correlation coefficient for each brand is marked in bold}

\scalebox{.8}{
\footnotesize
\begin{tabular}{|c| c| c c c c c c c c c c|}
\hline
\multirow{2}{*}{\textbf{Brand}} & \multirow{2}{*}{\textbf{Between}} & \multicolumn{10}{l|}{\textbf{   Pearson correlation coefficient with different latency periods} }\\ \cline{3-12} 
 &  & @1month &@2 month &@3 month &@4 month &@5 month &@6 month &@7 month &@8 month &@9 month &@10 month   \\ \hline
 \multirow{1}{*}{NOKIA}& SPS$\longleftrightarrow$Product sale & 0.353 & \textbf{0.793}& 0.134 & 0.324& 0.101 & 0.073 & -0.291 & 0.147& 0.117&0.097  \\ 
 
 & SPS-R $\longleftrightarrow$Product sale & 0.272& -0.297& \textbf{0.356} & 0.178 & -0.245 & 0.146 & 0.074 & -0.232 & -0.147&-0.103   \\ \cline{1-12}
 \multirow{2}{*}{SAMSUNG}&SPS$\longleftrightarrow$Product sale  & 0.781& \textbf{0.841} & 0.773 & 0.456 & 0.303 & 0.089 & -0.178 & -0.007 & 0.027& 0.012  \\
 
 & SPS-R$\longleftrightarrow$Product sale & -0.188& 0.163& 0.051 & -0.043& \textbf{0.203} & 0.068 & -0.179 & -0.172 & -0.265&-0.443   \\ \cline{1-12}
 \multirow{2}{*}{PHILIPS}& SPS$\longleftrightarrow$Product sale  & 0.380 & \textbf{0.728}& 0.658 & 0.658 & 0.363 & 0.296 & 0.182 & 0.419& -0.191&-0.232  \\ 
 
 & SPS-R $\longleftrightarrow$Product sale & -0.002 &-0.082& 0.158 & \textbf{0.429} & 0.138 & 0.221 & 0.305 & 0.326 & 0.169& -0.131   \\ \cline{1-12}
 \multirow{2}{*}{CANON}& SPS$\longleftrightarrow$Product sale  & 0.719 & 0.901& \textbf{0.929} & 0.701 & 0.651 & 0.647 & 0.538 & 0.487 & 0.423&0.485  \\
 
 & SPS-R $\longleftrightarrow$Product sale & 0.345 & \textbf{0.489}& 0.215 & 0.137 & 0.103 & 0.034 & 0.021 & 0.014 & -0.127&-0.0951   \\ \cline{1-12}
 \multirow{2}{*}{LENOVO}& SPS$\longleftrightarrow$Product sale  & \textbf{0.712} &0.613& 0.459 & 0.265 & 0.207 & -0.029 & -0.061 & -0.143 & -0.238&-0.207  \\ 
 & SPS-R $\longleftrightarrow$Product sale & \textbf{0.198} & 0.129& 0.066 & -0.41 & -0.152 & -0.034 & -0.029 & -0.192 & -0.281&-0.301 \\ \cline{1-12}

\end{tabular}
} 
\label{tab1}
\end{table*}
Here we measure financial health of a company in terms of monthly sales figure.
We make a basic assumption that number of new reviews is directly proportional to the sales figure.
We consider the total number of new reviews gained by all products of company in a month as its number of sales in that month.

\begin{figure}[t]

  \includegraphics[width=1\linewidth]{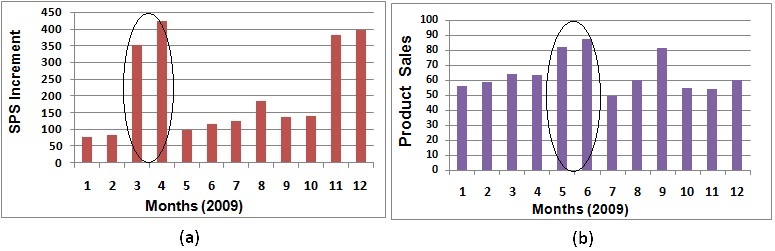}
  \caption{(a) Monthly $\sps$ increments of Nokia in 2009. (b)Monthly product sales of Nokia in 2009. Marked months show the short duration effects}
  \label{fig:sps2009}

  \centering
   \vspace{5mm}
  \includegraphics[width=1\linewidth]{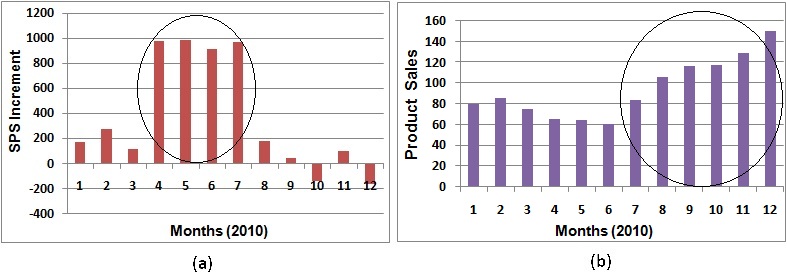}
  \caption{(a) Monthly $\sps$ increments of Nokia in 2010. (b)Monthly product sales of Nokia in 2010}
  \label{fig:sps2010}
  \vspace{5mm}
  \includegraphics[width=1\linewidth]{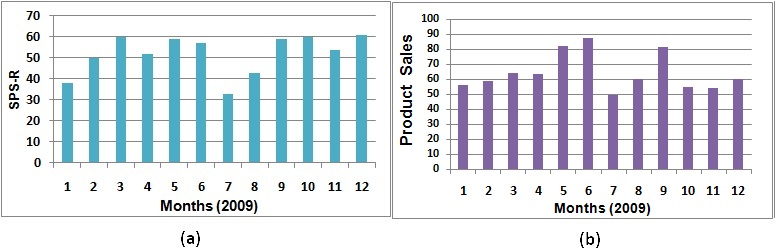}
  \caption{(a) Monthly $\base$ of Nokia in 2009. (b)Monthly product sales of Nokia in 2009}
  \label{fig:sps-r2009}
  \vspace{5mm}
  \includegraphics[width=1\linewidth]{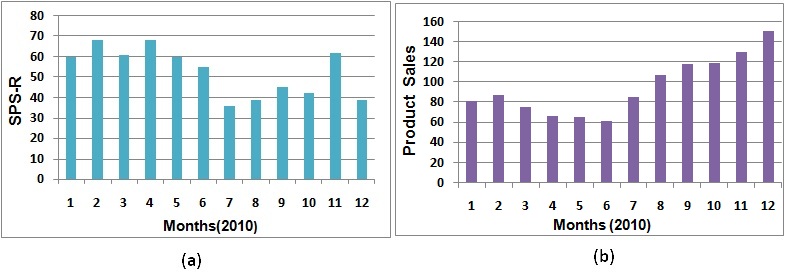}
  \caption{(a) Monthly $\base$ of Nokia in 2010. (b)Monthly product sales of Nokia in 2010}
  \label{fig:sps-r2010}

\end{figure}

Fig.~\ref{fig:sps2009}(a)  and ~\ref{fig:sps2009}(b) show the  relation between monthly increment of $\sps$ and monthly product sale of Nokia in 2009. 
In Fig.~\ref{fig:sps2009}(a) X-axis and Y-axis indicate the 12 months of 2009 and monthly increments of $\sps$ of Nokia respectively. 
In Fig.~\ref{fig:sps2009}(b) X-axis and Y-axis indicate the 12 months of 2009 and monthly product sales of Nokia respectively.
In Fig.~\ref{fig:sps2009}(a) $\sps$ increment of $3^{rd}$ (March) and $4^{th}$ (April) months are high compare to previous months and as a result in Fig.~\ref{fig:sps2009} (b) product sale of $5^{th}$ (May) and $6^{th}$ (June) months are high compare to previous months. 
In Fig.~\ref{fig:sps2009}(a) suddenly $\sps$ increment drops in $5^{th}$ month (May) and as a result in Fig.~\ref{fig:sps2009}(b) product sale is dropped in 7$^{th}$ month (July). 
Further from $6^{th}$ to $8^{th}$ month, $\sps$ increment gains and for this reason product sale is increased in the $8^{th}$ and $9^{th}$ month as shown in Fig.~\ref{fig:sps2009}. 
We can observe the same trends in later months as well.  

In $11^{th}$ and $12^{th}$ months of 2009, $\sps$ increment is highly increased (Fig.~\ref{fig:sps2009} (a)), as a result in $1^{st}$, $2^{nd}$ and $3^{rd}$ months of 2010 product sale goes upwards comparing to last three months as shown in Fig.~\ref{fig:sps2010}(b).

We have also observed Nokia's monthly $\base$ in the year of 2009 and 2010 as shown in Fig.~\ref{fig:sps-r2009} and \ref{fig:sps-r2010}. 
Fig.~\ref{fig:sps-r2009}(a)  and~\ref{fig:sps-r2009}(b) show the  relation between monthly $\base$ and monthly product sale of Nokia in 2009. 
From Fig.~\ref{fig:sps-r2009}  and Fig.~\ref{fig:sps-r2010}, it is very difficult to establish any significant positive  correlation between monthly $\base$ and monthly product sale of Nokia in 2009 and in later periods as well. 
So our proposed method $\sps$ is more effective than $\base$ to predict financial health of a brand.

\subsection{Comparison between $\sps$ and $\base$ }\label{subsec:comparison}
To evaluate our prediction of financial health and to establish our method's effectiveness  we correlate current monthly increment of $\sps$ with monthly sales figure from a future month. 
We compare our proposed method $\sps$ with a base line approach $\base$ , where $\base$ is based on the concept of Net Promoter Score($\nps$). 
Comparisons of Figs.~\ref{fig:sps2009} and \ref{fig:sps2010} with  Figs.~\ref{fig:sps-r2009} and ~\ref{fig:sps-r2010} clearly show the limitation of $\base$ and $\nps$.

To establish the limitation of $\base$ numerically we evaluate Pearson correlation coefficient between monthly $\sps$ increments and monthly sales figure over consecutive twelve months.
Consecutive twelve months of sales figures starts with a latency of few months with respect to that of $\sps$ increments to figure out the prediction horizon of $\sps$.
We evaluate Pearson correlation coefficient for different latency periods from one month up to ten months.

Table~\ref{tab1} shows correlation between increments of $\sps$ of some brands with their product sale. 
Furthermore,  Table~\ref{tab1} presents correlation between the base line approach $\base$ and product sale.
We consider monthly increments of $\sps$ and monthly $\base$ from the month of January to December in 2009 for aforementioned five brands. 

In Table~\ref{tab1}, the data in column named $@1 month$ means latency period is one month and it indicates that correlation coefficient is calculated between monthly increments of $\sps$ of a brand from January to December of 2009  with monthly product sales of that brand from February 2009 to January 2010.
The latency period which gives the maximum correlation is assumed to be the actual latency of $\sps$ to show its effects on financial health of the brand.
For example, Nokia gives maximum correlation coefficient ($0.793$) at latency of $2$ months, i.e., in case of Nokia the increment of its $\sps$ shows effects on product sales after two months. 

Similarly, we evaluate Pearson correlation coefficient between monthly $\base$ and monthly sales figure over consecutive twelve months for those five companies.
As shown in Table \ref{tab1}, our proposed method $\sps$ gives much better correlation with product sale than $\base$ for all brands. 

\subsection{Latency Periods}\label{subsec:latency}
In Table~\ref{tab2} we present correlation coefficients between increments of $\sps$ and product sale of different brands from different sectors. 
Latency period of a brand corresponds to the maximum correlation coefficient in the respective row which is written in bold fonts in Table~\ref{tab2}.
We compare latency periods of $\sps$ for different brands.
We observe that in sectors like electronic appliances and camera, latency periods have similar values between $2-3$ months.
However, in case of watch and laptop they disagree.
We have noticed that in general increment of $\sps$ reflects on sale in one to five months. 
\renewcommand{\arraystretch}{1.4}
\begin{table*} [t]
\centering
\addtolength{\tabcolsep}{.01pt}
%\centering
\caption{Pearson correlation coefficient between increments of $\sps$ and product sales of different brands from different sectors. Max correlation coefficient for each brand is marked in bold} 

\scalebox{.8}{
\footnotesize
\begin{tabular}{|c|c|c c c c c c c c c c|}
\cline{1-11}\hline
\multirow{1}{*}{\textbf{Sector}} &\multirow{1}{*}{\textbf{Brand}}  & \multicolumn{10}{l|}{\textbf{   Pearson correlation coefficient with different latency periods} }\\ \cline{3-12}
 
 &&@1month &@2month &@3 month &@4 month &@5 month &@6 month &@7 month &@8 month &@9 month &@10 month    \\ \cline{1-12}
 
 \multirow{1}{*}{Mobile}&\multirow{1}{*}{NOKIA}&  0.353 & \textbf{0.793}& 0.134 & 0.324& 0.101 & 0.073 & -0.291 & 0.147& 0.117&0.097  \\ \cline{1-12}\hline

 &\multirow{1}{*}{SAMSUNG} & 0.781& \textbf{0.841} & 0.773 & 0.456 & 0.303 & 0.089 & -0.178 & -0.007 & 0.027& 0.012  \\ %\cline{2-12} 
 
 &\multirow{1}{*}{LG}  & 0.332 & 0.289& \textbf{0.590} & 0.309 & 0.197 & -0.012 & -0.027 & 0.023 & 0.129&0.177  \\ %\cline{2-12}

 \multirow{1}{*}{Electronic Appliance}&\multirow{1}{*}{SONY} & 0.458 & \textbf{0.684}& 0.343 & 0.388 & 0.437 & 0.239 & 0.028 & 0.156 & -0.052&-0.031 \\ %\cline{2-12} 
 
 \multirow{1}{*}{}&\multirow{1}{*}{PHILIPS} & 0.380 & \textbf{0.728}& 0.658 & 0.658 & 0.363 & 0.296 & 0.182 & 0.419& -0.191&-0.232  \\ %\cline{2-12}
 
 &\multirow{1}{*}{PANASONIC} & 0.165 &0.233& 0.277 & 0.196 & 0.322 & 0.397 & \textbf{0.607}& 0.467& 0.233 &0.041 \\ \cline{2-12} \hline

&\multirow{1}{*}{CANON} & 0.719 & 0.901& \textbf{0.929} & 0.701 & 0.651 & 0.647 & 0.538 & 0.487 & 0.423&0.485  \\ %\cline{2-12}
 
 \multirow{1}{*}{Camera} &\multirow{1}{*}{NIKON} &0.418  & \textbf{0.737}& 0.517 &0.035 &0.017 &0.013  &0.018 &-0.172  &-0.017 &-0.054  \\ %\cline{2-12}
 
 & \multirow{1}{*}{KODAK} &0.341  & \textbf{0.719}&0.127  &0.197  &0.231  &0.292  &0.332  &0.271  &0.291 & -0.131 \\ \cline{2-12}\hline

  &\multirow{1}{*}{TITAN} &0.028  &\textbf{0.776}&0.352  & 0.279& -0.092& -0.014  &-0.174& -0.201 &0.189&0.105   \\ %\cline{2-12}
  
   \multirow{1}{*}{Watch}&\multirow{1}{*}{TIMAX} &-0.232  &-0.273 &-0.288  &0.219 &\textbf{0.867} &0.516  &0.231 &0.063  &0.145 &-0.026  \\ %\cline{2-12} 
  
  &\multirow{1}{*}{SONATA}&0.167  &0.283 &0.182  &\textbf{0.749} & 0.329&0.197  &0.061 &-0.167  &-0.029 &-0.277 \\ \cline{2-12}\hline

 \multirow{1}{*}{Laptop} &\multirow{1}{*}{LENOVO} & \textbf{0.712} &0.613& 0.459 & 0.265 & 0.207 & -0.029 & -0.061 & -0.143 & -0.238&-0.207   \\ %\cline{2-12}

 &\multirow{1}{*}{ASUS}&  0.219 &0.391 &\textbf{0.612} & 0.261 & 0.181 & 0.092& 0.176 &0.102  &-0.032 & -0.172\\ \cline{2-12} \hline

 \multirow{1}{*}{Sports wear} &\multirow{1}{*}{ADIDAS} &-0.191  & 0.219&0.321  &0.279 &\textbf{0.551} &0.178  &0.023 &0.160  &0.069 &-0.159  \\ %\cline{2-12}
 
 \multirow{1}{*}{}&\multirow{1}{*}{NIKE} &0.358  &0.310 & 0.281 &\textbf{0.674}& 0.451&0.092 &0.167  &-0.128 &0.152  &0.0391  \\ 
 
 \multirow{1}{*}{}&\multirow{1}{*}{ REEBOK} &0.156  &0.456 & 0.487 &\textbf{0.506}& 0.320&0.103 &0.067  &-0.104 &-0.188  &0.0244  \\ \cline{2-12}\hline 

\end{tabular}

}
\label{tab2}
\end{table*}

\subsection{Long and Short Duration Effects}\label{subsec:duration}
We observe that prolong consecutive high or low increments of $\sps$ has different effects on product sale than a short duration of high or low increments of $\sps$. 
Fig.~\ref{fig:sps2009} shows short duration effect of Nokia in 2009. 
 In Fig.~\ref{fig:sps2009}(a) increment of $\sps$ in $3^{rd}$ (March) and $4^{th}$ (April) months are high compare to previous months and as a result in Fig.~\ref{fig:sps2009}(b) product sale of $5^{th}$ (May) and $6^{th}$ (June) months are high compare to previous month. 
Duration of high increments of $\sps$ is for only two months and as a result duration of high product sale is also short (two months). 
This kind of outcome of $\sps$ increment is called short duration effect. 

On the other hand, Fig.~\ref{fig:sps2010} shows long duration effect of $\sps$ on Nokia in 2010.
In Fig.~\ref{fig:sps2010}(a) increments of $\sps$  in $4^{th}$ to $7^{th}$  (April to July) are high compare to previous months and as a result in Fig.~\ref{fig:sps2010}(b) product sales of $7^{th}$ to $12^{th}$ (July to December) months are high compare to previous months. 
Here scenario is quite different. 
The duration of $\sps$ increment is  for four months and as a result duration of high product sale is also for longer time period (six months). 
It is called long duration effect and it is more beneficial for company's financial growth than short duration effect.

\begin{figure}[t]
  \vspace{5mm}
  \includegraphics[width=1\linewidth]{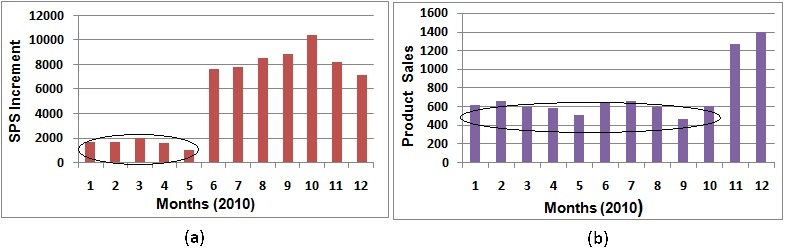}
  \caption{Long duration effects on Sony in 2010}
  \label{fig:longeffsony}
  \vspace{5mm}

  \includegraphics[width=1\linewidth]{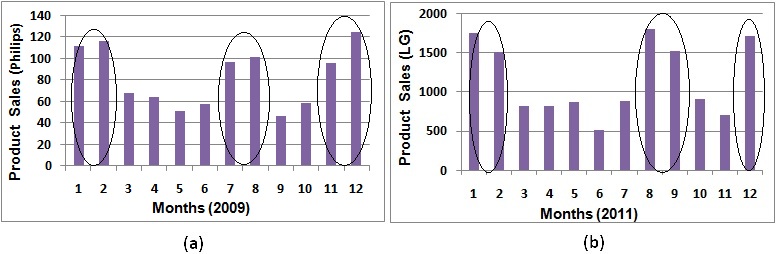}
  \caption{Exclusive Offer effects on Philips and LG}
  \label{fig:offerLG}
\end{figure}

 Fig.~\ref{fig:longeffsony}(a) shows that Sony's $\sps$ increment are very low from  $1^{st}$ to $5^{th}$  (January to May) months and as a result in Fig.~\ref{fig:longeffsony}(b) product sales are also low from  $2^{nd}$ to $10^{th}$ (February to October) months. 
The duration of low increments of $\sps$  for five months results a longer duration (nine months) of poor product sale. 
So long duration effect is also present in case of poor performance of the brand.  

Companies should focus on effective positive long duration effect to enhance their financial health by applying different incentive schemes specifically for influential customers and establish strategy to encourage the detractors to purchase products again by enhancing the quality of products or by more aggressive advertisements/ incentives or both.
At the same time companies should be alarmed by any long duration of low or negative increments of $\sps$. 

Usually brands promote special seasonal offers like exclusive summer sale, Christmas sale to attract customers. 
Fig.~\ref{fig:offerLG} shows the effects of those exclusive offers on monthly product sales of Philips and LG brands.
It shows that product sales in December, January (Christmas season) and July, August, September (Summer season) is high compare to other months.

\section{Conclusion and Future works}\label{sec:conclusion}
In this paper, we propose  a method ($\sps$) and a tool (\textit{review network}) for predicting financial health of an online shopping brand. 
Our proposed method and tool are unbiased because there is no manual intervention. 
We show that not only customers' reviews and ratings but also their centrality in review network has a good impact in predicting market growth of a brand. 
Helpfulness score is used for validation of customers' reviews and ratings.
Helpfulness score has a great impact on $\sps$ of any brand.
We have noticed that in general prediction from SPS reflects on product sale in one to five months.
Loyal customers with high centrality  and detractors are identified by the brand more accurately using our method.
Company can decides different incentive schemes for specially influential customers and establish strategy to encourage the detractors to purchase products again by enhancing the quality of products or by more aggressive advertisements/ incentives or both.

There are several interesting directions that need further investigations. 
First, in this work, we consider centrality of customers only in their review network and we would like to further investigate centrality of customers in their social networks. 
Sometimes customers share the experiences about their purchase products in their social networks and their friend circle give reply on these posts. 
As for example a promoter with 100 social friends in her social network and a promoter with 5 friends can not have the same effects on brand's growth. 
We have applied our proposed method only on Amazon.com online review data set. 
Further we would like to use data sets of other online merchandise sites, online hotel booking sites etc. 

In this work we make a simple assumption about product sales figure of a brand from its number of reviews.
It will be more interesting and accurate to consider actual sales figures of the brand.
There will be a latency period which gives the time required for showing the effect of SPS on actual growth of the firm. 
This latency period depends on data, season, company offers, marketing strategy etc. 
We would like to investigate if these parameters can be included in our method that may provide more accurate prediction.

In this paper we evaluate absolute $\sps$ increment value of a brand at any time instance.
The actual metric based on specific threshold value of $\sps$ increment will give more clear indication of the performance of a brand.
Modeling that metric will be an interesting research topic.
We will try to apply this concept in recommender system and recommend item to user based on $SPS$.

\bibliographystyle{IEEEtran}

%\bibliography{supriyo}

\begin{thebibliography}{10}
\providecommand{\url}[1]{#1}
\csname url@samestyle\endcsname
\providecommand{\newblock}{\relax}
\providecommand{\bibinfo}[2]{#2}
\providecommand{\BIBentrySTDinterwordspacing}{\spaceskip=0pt\relax}
\providecommand{\BIBentryALTinterwordstretchfactor}{4}
\providecommand{\BIBentryALTinterwordspacing}{\spaceskip=\fontdimen2\font plus
\BIBentryALTinterwordstretchfactor\fontdimen3\font minus
  \fontdimen4\font\relax}
\providecommand{\BIBforeignlanguage}[2]{{%
\expandafter\ifx\csname l@#1\endcsname\relax
\typeout{** WARNING: IEEEtran.bst: No hyphenation pattern has been}%
\typeout{** loaded for the language `#1'. Using the pattern for}%
\typeout{** the default language instead.}%
\else
\language=\csname l@#1\endcsname
\fi
#2}}
\providecommand{\BIBdecl}{\relax}
\BIBdecl

\bibitem{NPS}
F.~F. Reichheld, ``The one number you need to grow,'' \emph{Harvard business
  review}, vol.~81, no.~12, pp. 46--55, 2003.

\bibitem{Rafiei2011}
F.~M. Rafiei, S.~Manzari, and S.~Bostanian, ``Financial health prediction
  models using artificial neural networks, genetic algorithm and multivariate
  discriminant analysis: Iranian evidence,'' \emph{Expert Systems with
  Applications}, vol.~38, no.~8, pp. 10\,210--10\,217, 2011.

\bibitem{huang2014}
R.~Huang and E.~Sarig{\"o}ll{\"u}, ``How brand awareness relates to market
  outcome, brand equity, and the marketing mix,'' in \emph{Fashion Branding and
  Consumer Behaviors}.\hskip 1em plus 0.5em minus 0.4em\relax Springer, 2014,
  pp. 113--132.

\bibitem{oliver2014}
R.~L. Oliver, \emph{Satisfaction: A behavioral perspective on the
  consumer}.\hskip 1em plus 0.5em minus 0.4em\relax Routledge, 2014.

\bibitem{aaker2012}
D.~A. Aaker, \emph{Building strong brands}.\hskip 1em plus 0.5em minus
  0.4em\relax Simon and Schuster, 2012.

\bibitem{anderson1994}
E.~W. Anderson, C.~Fornell, and D.~R. Lehmann, ``Customer satisfaction, market
  share, and profitability: Findings from sweden,'' \emph{The Journal of
  marketing}, pp. 53--66, 1994.

\bibitem{reichheld1990}
F.~F. Reichheld and J.~W. Sasser, ``Zero defections: Quality comes to
  services.'' \emph{Harvard business review}, vol.~68, no.~5, pp. 105--111,
  1990.

\bibitem{anderson1993}
E.~W. Anderson and M.~W. Sullivan, ``The antecedents and consequences of
  customer satisfaction for firms,'' \emph{Marketing science}, vol.~12, no.~2,
  pp. 125--143, 1993.

\bibitem{boulding1993}
W.~Boulding, A.~Kalra, R.~Staelin, and V.~A. Zeithaml, ``A dynamic process
  model of service quality: from expectations to behavioral intentions,''
  \emph{Journal of marketing research}, vol.~30, no.~1, p.~7, 1993.

\bibitem{johnson1991}
M.~D. Johnson and C.~Fornell, ``A framework for comparing customer satisfaction
  across individuals and product categories,'' \emph{Journal of economic
  psychology}, vol.~12, no.~2, pp. 267--286, 1991.

\bibitem{kaplan2016}
J.~Kaplan, ``The inventor of customer satisfaction surveys is sick of them,
  too,'' \emph{Bloomberg Technology}, 2016.

\bibitem{WOM}
\BIBentryALTinterwordspacing
R.~V. Kozinets, K.~de~Valck, A.~C. Wojnicki, and S.~J. Wilner, ``Networked
  narratives: Understanding word-of-mouth marketing in online communities,''
  \emph{Journal of Marketing}, vol.~74, no.~2, pp. 71--89, 2010. [Online].
  Available: \url{https://doi.org/10.1509/jmkg.74.2.71}
\BIBentrySTDinterwordspacing

\bibitem{grisaffe2007}
D.~B. Grisaffe, ``Questions about the ultimate question: conceptual
  considerations in evaluating reichheld's net promoter score (nps),''
  \emph{Journal of Consumer Satisfaction, Dissatisfaction and Complaining
  Behavior}, vol.~20, p.~36, 2007.

\bibitem{raassens2017}
N.~Raassens and H.~Haans, ``Nps and online wom: Investigating the relationship
  between customers’ promoter scores and ewom behavior,'' \emph{Journal of
  Service Research}, p. 1094670517696965, 2017.

\bibitem{brand}
K.~Zhang, D.~Downey, Z.~Chen, Y.~Xie, Y.~Cheng, A.~Agrawal, W.-k. Liao, and
  A.~Choudhary, ``A probabilistic graphical model for brand reputation
  assessment in social networks,'' in \emph{Proceedings of the 2013 IEEE/ACM
  International Conference on Advances in Social Networks Analysis and
  Mining}.\hskip 1em plus 0.5em minus 0.4em\relax ACM, 2013, pp. 223--230.

\bibitem{sentiment1}
S.~V. Wawre and S.~N. Deshmukh, ``Sentiment classification using machine
  learning techniques,'' \emph{Int. J. Sci. Res}, vol.~5, no.~4, pp. 1--3,
  2016.

\bibitem{sentiment2}
E.~Riloff and J.~Wiebe, ``Learning extraction patterns for subjective
  expressions,'' in \emph{Proceedings of the 2003 conference on Empirical
  methods in natural language processing}.\hskip 1em plus 0.5em minus
  0.4em\relax Association for Computational Linguistics, 2003, pp. 105--112.

\bibitem{review1}
J.~Liu and S.~Seneff, ``Review sentiment scoring via a parse-and-paraphrase
  paradigm,'' in \emph{Proceedings of the 2009 Conference on Empirical Methods
  in Natural Language Processing: Volume 1-Volume 1}.\hskip 1em plus 0.5em
  minus 0.4em\relax Association for Computational Linguistics, 2009, pp.
  161--169.

\bibitem{review2}
R.~McDonald, K.~Hannan, T.~Neylon, M.~Wells, and J.~Reynar, ``Structured models
  for fine-to-coarse sentiment analysis,'' in \emph{Annual meeting-association
  for computational linguistics}, vol.~45, no.~1, 2007, p. 432.

\bibitem{socialmax}
S.~Bhagat, A.~Goyal, and L.~V. Lakshmanan, ``Maximizing product adoption in
  social networks,'' in \emph{Proceedings of the fifth ACM international
  conference on Web search and data mining}.\hskip 1em plus 0.5em minus
  0.4em\relax ACM, 2012, pp. 603--612.

\bibitem{socialmax1}
S.~Lei, S.~Maniu, L.~Mo, R.~Cheng, and P.~Senellart, ``Online influence
  maximization,'' in \emph{Proceedings of the 21th ACM SIGKDD International
  Conference on Knowledge Discovery and Data Mining}.\hskip 1em plus 0.5em
  minus 0.4em\relax ACM, 2015, pp. 645--654.

\bibitem{socialmax2}
M.~Zhang, C.~Dai, C.~Ding, and E.~Chen, ``Probabilistic solutions of influence
  propagation on social networks,'' in \emph{Proceedings of the 22nd ACM
  international conference on Information \& Knowledge Management}.\hskip 1em
  plus 0.5em minus 0.4em\relax ACM, 2013, pp. 429--438.

\bibitem{socialmax3}
J.~Zhang, C.~Wang, and J.~Wang, ``Learning temporal dynamics of behavior
  propagation in social networks.'' in \emph{AAAI}, 2014, pp. 229--236.

\bibitem{shen2011}
Z.~Shen and N.~Sundaresan, ``ebay: an e-commerce marketplace as a complex
  network,'' in \emph{Proceedings of the fourth ACM international conference on
  Web search and data mining}, 2011, pp. 655--664.

\bibitem{fackorloyal}
H.~Xu, D.~Liu, H.~Wang, and A.~Stavrou, ``E-commerce reputation manipulation:
  The emergence of reputation-escalation-as-a-service,'' in \emph{Proceedings
  of the 24th International Conference on World Wide Web}.\hskip 1em plus 0.5em
  minus 0.4em\relax International World Wide Web Conferences Steering
  Committee, 2015, pp. 1296--1306.

\bibitem{eigenvector}
L.~Sol{\'a}, M.~Romance, R.~Criado, J.~Flores, A.~Garc{\'\i}a~del Amo, and
  S.~Boccaletti, ``Eigenvector centrality of nodes in multiplex networks,''
  \emph{Chaos: An Interdisciplinary Journal of Nonlinear Science}, vol.~23,
  no.~3, p. 033131, 2013.

\bibitem{betweenness}
A.~Mahmoody, C.~E. Tsourakakis, and E.~Upfal, ``Scalable betweenness centrality
  maximization via sampling,'' \emph{arXiv preprint arXiv:1609.00790}, 2016.

\bibitem{closeness}
Y.~Du, C.~Gao, X.~Chen, Y.~Hu, R.~Sadiq, and Y.~Deng, ``A new closeness
  centrality measure via effective distance in complex networks,'' \emph{Chaos:
  An Interdisciplinary Journal of Nonlinear Science}, vol.~25, no.~3, p.
  033112, 2015.

\bibitem{he2016}
R.~He and J.~McAuley, ``Ups and downs: Modeling the visual evolution of fashion
  trends with one-class collaborative filtering,'' in \emph{Proceedings of the
  25th International Conference on World Wide Web}, 2016, pp. 507--517.

\bibitem{mcauley2015}
J.~McAuley, C.~Targett, Q.~Shi, and A.~Van Den~Hengel, ``Image-based
  recommendations on styles and substitutes,'' in \emph{Proceedings of the 38th
  International ACM SIGIR Conference on Research and Development in Information
  Retrieval}, 2015, pp. 43--52.

\end{thebibliography}
\end{document}